\documentclass[a4paper,fleqn,usenatbib]{mnras}

\usepackage{newtxtext,newtxmath}
\usepackage[T1]{fontenc}
\usepackage{ae,aecompl}
\usepackage{graphicx}	% Including figure files
\usepackage{amsmath}	% Advanced maths commands
\usepackage{amssymb}	% Extra maths symbols

\title[On the assumption of strong lensing]%
  {Can continuous density profile correctly describe strong gravitational lensing?}

\author[Tian]{Shuxun Tian\thanks{E-mail: tshuxun@whu.edu.cn}\\%
  School of Physics and Technology, Wuhan University, Wuhan 430072, China\\%
  Center for Cosmology and Gravitational Waves, Wuhan University, Wuhan 430072, China}

\date{Accepted XXX. Received YYY; in original form ZZZ}

\pubyear{2017}

\begin{document}
\label{firstpage}
\pagerange{\pageref{firstpage}--\pageref{lastpage}}
\maketitle

\begin{abstract}
  Strong gravitational lensing is an important tool to probe the universe. In the theoretical analysis of gravitational lensing, it is assumed that continuous density profile can correctly describe the lens galaxy. But in fact this assumption has never been rigorously tested. In this paper, we discuss this issue, and point out that if we use discrete density profile to model the lens galaxy, then the position of the images does not change, but the magnification will be increased. Strongly lensed gravitational waves could test this conclusion in the future.
\end{abstract}


\section{Introduction}
Strong gravitational lensing is a very important astronomical phenomenon, which can be used to study many fundamental physical problems, such as dark matter \citep{Kamada2016,Parry2016} and cosmology \citep{Cao2015,Linder2016,Refsdal1964}. \cite{Walsh1979} reported the first strong lensing event. With the launch of several survey projects \citep[e.g.][]{Auger2009,Bolton2008,Brownstein2012}, now people have found hundreds of events. At the same time, theoretical analysis of gravitational lensing has also been in-depth development. Typical models include singular isothermal sphere (SIS) model, elliptical galaxy model, NFW density profile model, etc \citep[see][for reviews]{Bartelmann2010,Narayan1996}.

In the theoretical analysis of gravitational lensing, people always assume the continuous density profile can correctly describe the lens galaxy, even though this assumption has never been rigorously tested. One must admit that the density profile of galaxies is not continuous. The density distribution can be described by a Dirac $\delta$ function around the black hole, neutron star, or even a general stellar. Discrete density profile inevitably leads to a deviation of the gravitational potential on small scales from the continuous case. As we will see, if we use discrete density profile to model the lens, then the position of images remains unchanged, but the magnification will be increased.

In this paper, we first calculate the influence of a compact object on the original image\footnote{Original image refers to the image formed by the continuous density profile.} where the lens galaxy is described by the SIS model. The result shows this compact object will increase the brightness of the original image. This is a trivial result, but raise a serious problem: can continuous density profile correctly describe the strong gravitational lensing? Our result gives a negative answer. In observations, we point out strongly lensed gravitational waves can verify our conclusion.

\section{Influence of a point mass on the SIS model}
\begin{figure}
	\includegraphics[width=\columnwidth]{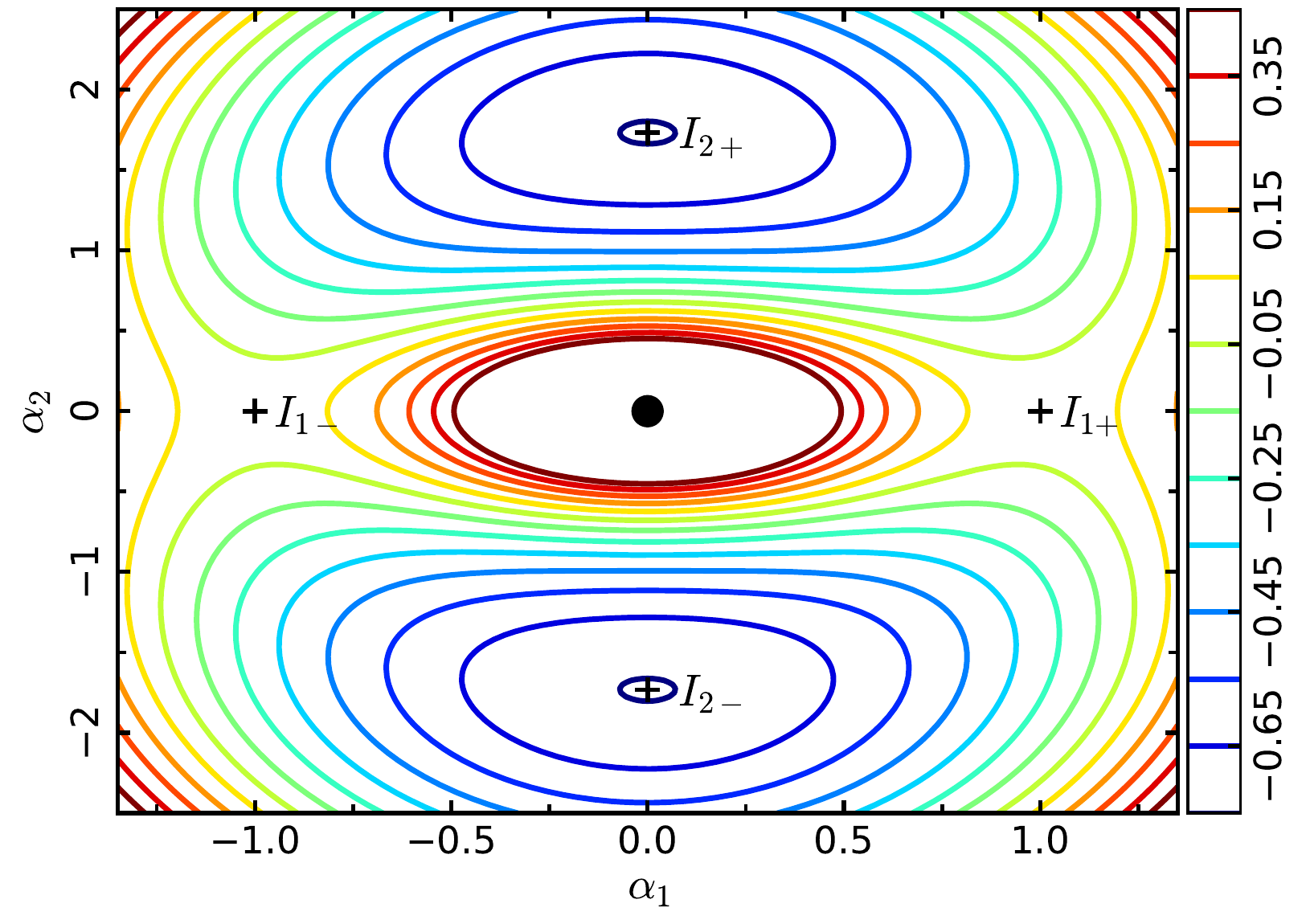}
    \caption{Contour plot of the time-delay function. The coordinates $\alpha_1$ and $\alpha_2$ are defined as $\vec{\theta}=(\beta+\theta_{E1}+\alpha_1\cdot\theta_{E2},\ \alpha_2\cdot\theta_{E2})$. We shift $t(\vec{\theta})$ to make $t(I_{1+})=0$, and set $M=50M_\odot$ (solar mass) to plot the colorbar. The unit of the colorbar is ms. The central black dot represents a compact object.}
    \label{fig:1}
\end{figure}

For a gravitational lensing system, we assume the lens galaxy is described by the SIS model with Einstein radius $\theta_{E1}$ and the source in the $\vec{\beta}$ direction. If $\beta<\theta_{E1}$, then two images appears, denote as $I_\pm$. The position of the images is $\theta_\pm=\beta\pm\theta_{E1}$, and the magnification $\mu_\pm=(1\mp\theta_{E1}/\theta_\pm)^{-1}$. Assuming a compact object with mass of $M$ appears in the position of $\vec{\theta}_M$ in the lens galaxy. This compact object can be regard as a point mass. Since the Newtonian gravitational potential can be linearly superimposed, the total effective lensing potential can be written as
\begin{align}
  \psi(\vec{\theta})=\frac{D_{ls}}{D_s}\frac{4\pi\sigma_v^2}{c^2}|\vec{\theta}|+
  \frac{D_{ls}}{D_lD_s}\frac{4GM}{c^2}\ln|\vec{\theta}-\vec{\theta}_{M}|,
\end{align}
where $\sigma_v$ is the velocity dispersion. Based on the previous statement, we know $\theta_{E1}=4\pi\sigma_v^2D_{ls}/c^2D_s$. We denote $\theta_{E2}=\sqrt{4GMD_{ls}/c^2D_lD_s}$ hereafter. We can use Fermat's principle to determine the position of the images. Considering the time-delay function
\begin{equation}\label{eq:02}
  t(\vec{\theta})=\frac{(1+z_l)}{c}\frac{D_lD_s}{D_{ls}}\left[\frac{1}{2}(\vec{\theta}-\vec{\beta})^2-\psi(\vec{\theta})\right],
\end{equation}
then the images satisfy the condition $\nabla_{\vec{\theta}}\ t=0$. In the following, we assume $z_l=0.4$, $z_s=1$, $\sigma_v=220{\rm km}/{\rm s}$, and $\beta=\theta_{E1}/2$ as a typical lensing system. We also assume a flat universe with $H_0=70{\rm km/s/Mpc}$ and $\Omega_m=0.3$. In addition, we assume the compact object appears at the position of $I_+$. Intuitively, this is very occasional. But note that we only considered one compact object, while there are many in the real galaxy. According to the later analysis of event rate, it is ubiquitous that a compact object appears near the original image. Figure \ref{fig:1} plot the time-delay function for the typical lensing system, which shows a point mass splits the original image into four images. We denote these four images as $I_{1\pm}$ and $I_{2\pm}$, respectively. We will calculate the position of these images, the arrival time differences and the magnifications.

Solving $\nabla_{\vec{\theta}}\ t=0$ gives
\begin{align}
  \vec{\theta}(I_{1\pm})&=(\beta+\theta_{E1}\pm\theta_{E2},\ 0),\\
  \vec{\theta}(I_{2\pm})&=(\beta+\theta_{E1}-\frac{\theta_{E2}^2}{2\beta},\ \theta_2),
\end{align}
where $\theta_2$ satisfy the equation
\begin{equation}
  1-\frac{\theta_{E1}}{\sqrt{\left(\beta+\theta_{E1}-\displaystyle\frac{\theta_{E2}^2}{2\beta}\right)^2+\theta_2^2}}
  -\frac{\theta_{E2}^2}{\displaystyle\frac{\theta_{E2}^4}{4\beta^2}+\theta_2^2}=0.
\end{equation}
We solve the above equation numerically. After obtaining the position of the images, we can use equation (\ref{eq:02}) to calculate the arrival time difference between the images. Symmetry gives $t(I_{2+})=t(I_{2-})$. Numerical result gives $t(I_{1+})\approx t(I_{1-})$, and
\begin{equation}\label{eq:06}
  t(I_{1+})-t(I_{2+})\approx1.5\frac{M}{100M_{\odot}}{\rm ms}.
\end{equation}
The above result is very accurate when $M<100M_\odot$ (the relative error is much less than $1\%$). Now we calculate the magnification of the images. Considering the inverse magnification tensor $\mathcal{A}$, which is defined as
\begin{equation}
  \mathcal{A}_{ij}=\delta_{ij}-\frac{\partial^2\psi}{\partial\theta_i\partial\theta_j},
\end{equation}
then the magnification $\mu=1/|{\rm det\ }\mathcal{A}|$. Numerical result gives $\mu(I_{1\pm})\approx0.75$, and $\mu(I_{2\pm})\approx2.25$. Likewise, this result is also very accurate when $M<100M_\odot$. More importantly, the magnification of the images is independent of the central star's mass. In the previous parameter settings, SIS model gives $\mu(I_+)=3$, while the total magnification of the four new images is $\mu=6$. So the compact object that appears near the original image will increase its brightness. This is a trivial result, just as traditional gravitational lensing can increase the brightness of the source. But once we realized the galaxy contains numerous compact objects, this calculation raise a serious problem: can continuous density profile correctly describe strong gravitational lensing? If we model the lens density in a discrete way, can we obtain the same results as the continuous case?

For the point source in the lensing system, the original image will be affected if the compact object in the lens galaxy appears near it. Here we estimate the probability of such an event. As mentioned before, the magnification of the images is independent of the central star's mass. Therefore, a solar--mass stellar can produce observable effects. The lens galaxy is described by the SIS model with surface mass density $\Sigma(\xi)=\sigma_v^2/2G\xi$. Assuming the mass of all the compact objects in the galaxy is $M$, then its surface number density $f(\xi)=\Sigma(\xi)/M$. We can assume the original image will be affected if $|\vec{\theta}_M-\vec{\theta}(I_+)|<\theta_{E2}$. This assumption is reasonable, as the Einstein radius represents the characteristic length of the gravitational potential. Then the average number of compact objects that can affect the original image is
\begin{equation}\label{eq:08}
  \bar{N}=f(\xi)\cdot\pi(D_l\theta_{E2})^2=0.5.
\end{equation}
We use $\xi=D_l\theta_{E1}$ in the last equality. This result means almost half of the strongly lensed images of point sources are affected by the compact objects in the lens galaxy. If the mass of the compact object is small, the corresponding arrival time difference is unobservable, but the total magnification is observable. If the source of the lensing system is an extended source, such as a galaxy, the effect of the compact objects should be an averaged result. It can be seen from previous calculation that the point mass in the lens galaxy hardly affects the position of the original image. This phenomenon can also be attributed to that the position of the image depend only on the large--scale gravitational potential of the lens galaxy. But the magnification is different, which is closely related to the small--scale gravitational potential. We can regard the source galaxy as a set of point source, then equation (\ref{eq:08}) indicates that half of the images will become brighter because of the influence of the compact objects in the lens galaxy. On the whole, the image of the source galaxy will become brighter. Thus, using discrete density profile to model the lens galaxy gives a greater magnification of the images.

\section{Astronomical test}
In principle, we have two ways to detect the influence of compact objects on strong gravitational lensing. One is to measure the arrival time difference of burst signals. \cite{Munoz2016} discussed the possibility of using fast radio bursts to detect massive compact halo objects. Their core idea is that the lensed fast radio bursts behave as bimodal, and observations can distinguish these two peaks if the arrival time difference reaches 1ms. Likewise, we can use the doubly peaked fast radio bursts to detect the effects of the point mass in the lens galaxy. But this method has a drawback that mass needs to be large. Equation (\ref{eq:06}) shows $66M_\odot$ is required for the compact object in order to produce a time difference of 1ms. And we still do not know how much of this kind of objects exist in the galaxy. Thus this method is not realistic.

In the other way, we can measure the magnification to detect the influence of compact objects on strong gravitational lensing. This method is not feasible in traditional optical observations because we do not know the intrinsic luminosity of the source galaxy. But gravitational waves make this method possible. Now people have observed five gravitational wave events, which opens the era of gravitational wave astronomy \citep{Abbott2016a,Abbott2016b,Abbott2017a,Abbott2017b,Abbott2017c}. The third generation of gravitational wave detector Einstein telescope are expected to observe 50--100 strongly lensed gravitational waves per year \citep{Biesiada2014}. For the lensed gravitational waves, the amplitude changes and the frequency remains unchanged. Note that under previous parameter settings, the physical length corresponding to the Einstein radius of a solar--mass stellar is 0.01pc, which is much larger than the wavelength of gravitational waves. So we can use optical approximation to deal with the motion of gravitational waves. In addition, the typical period of gravitational waves is much larger than the arrival time difference between different images. Thus, we can simply add the magnification of each image as the total magnification of the observed gravitational wave signal without regard to the effect of interference cancellation. If optical observations give the redshift of source galaxy, we can calculate the luminosity distance with a cosmology model. And we can obtain the chirp mass from the gravitational wave signal's phasing, i.e. the frequency information. Then we can calculate the intrinsic amplitude of the gravitational wave signal from the above quantities, and the magnification can be obtained from comparing with the observed amplitude. Optical observations can reconstruct the continuous density and gravitational potential profile of the lens galaxy, and then the theoretical magnification of the gravitational wave signal can be calculated. By comparing these two magnifications, we can determine whether the compact objects in the lens galaxy affect the strong gravitational lensing, more precisely, whether the continuous density profile gives the correct magnification.

\section{Conclusions}
In this paper, we discuss the issue that whether the continuous density profile can correctly describe strong gravitational lensing. We calculate the influence of a point mass on the SIS model, and find that the point mass splits the original image into four images while the total magnification doubled. This result implies that if the lens galaxy is modeled as a discrete set of point mass, the position of the gravitational lensing image does not change, but the magnification of each image will be greater than it in the continuous case. We point out strongly lensed gravitational waves can be used to measure the gravitational lensing magnification directly, and then it can determine whether the continuous density profile gives the correct magnification.

\section*{Acknowledgements}
This work was supported by the National Natural Science Foundation of China under grant No. 11633001.

% Don't change these lines
\bsp	% typesetting comment
\label{lastpage}
\end{document}